\documentclass[aps,pre,twocolumn,showpacs,floatfix,supperscriptaddress]{revtex4-1}
\usepackage{graphicx,bm,amsmath,dcolumn,longtable,color}
\usepackage{geometry,hyperref}
\graphicspath{{figures}}
\begin{document}
\title{Ising-like phase transition of an $n$-component Eulerian face-cubic model}
 \author{Chengxiang Ding}
\email{dingcx@ahut.edu.cn}
\affiliation{Department of Applied Physics, Anhui University of Technology, Maanshan 243002, China }
\author{Wenan Guo}
\email{waguo@bnu.edu.cn}
\affiliation{Physics Department, Beijing Normal University, Beijing 100875, China}
\author{Youjin Deng}
\email{yjdeng@ustc.edu.cn}
\affiliation{Hefei National Laboratory for Physical Sciences at Microscale, 
Department of Modern Physics, University of Science and Technology of China, Hefei, 230027, China}
\date{\today} 
\begin{abstract}
By means of Monte Carlo simulations and a finite-size scaling analysis,
we find a critical line of an $n$-component Eulerian face-cubic 
model on the square lattice and the simple cubic lattice in the region $v>1$, where $v$ is the bond weight. 
The phase transition belongs to the Ising universality class independent of $n$. 
The critical properties of the phase transition can also be captured by the percolation of the complement
of the Eulerian graph.

\end{abstract}
\pacs{05.50.+q, 64.60.Cn, 64.60.Fr, 75.10.Hk}
\maketitle 
\section{Introduction} 
In this paper, we study an $n$-component Eulerian face-cubic (EFC)
model\cite{henkERC, GuoERC} with partition sum
\begin{eqnarray}
Z=\sum\limits_G v^{N_b}n^{N_c}\, ,\label{fcb}
\end{eqnarray}
where $G$ is an Eulerian graph with $N_b$ bonds and $N_c$ clusters.
 The word `Eulerian' means the number of bonds connected to each site must be even.
A cluster is defined as an isolated site or a group of sites connected by the bonds.
This model originates from the face-cubic model\cite{facecub}, of which the Hamiltonian is 
\begin{eqnarray}
{\mathcal H}/k_BT=-\sum\limits_{<i,j>}[K\vec{s_i}\cdot\vec{s_j}+M(\vec{s_i}\cdot\vec{s_j})^2]. \label{hfcb}
\end{eqnarray}
Here $\vec{s_i}=(s_{i1},s_{i2},\cdots,s_{in})$ is an $n$-component cubic spin, with one and only one of the components has a nonzero value $\pm 1$. 
This model obviously combines the degrees of freedom of the Ising model and the $n$-state Potts model, thus the Hamiltonian can be alternatively written
as 
\begin{eqnarray}
{\mathcal H}/k_BT=-\sum\limits_{<i,j>}(K\sigma_i\sigma_j+M)\delta_{\tau_i\tau_j}, \label{hfcb2}
\end{eqnarray}
where $\sigma_i$ and $\tau_i$ are the Ising spin and the Potts spin, respectively. 
The Hamiltonian (\ref{fcb}) can be obtained by 
an graphical expansion of Hamiltonian (\ref{hfcb2}) under a restriction $\cosh K=\exp(-M)$, see Refs. \cite{henkERC, GuoERC} for details.
In the expansion, the bond weight $v$ is related to the inverse temperature $K$ by $v=\tanh K$. However, 
in the EFC model (\ref{fcb}) the bond weight $v$ is allowed to be larger than 1, and $n$ can be any real value, instead of an integer.

The face-cubic model has many applications, such as the adsorbed monolayers\cite{fcapp1}, 
the long polymer chains\cite{fcapp2,fcapp3} and so forth. 
The model has rich critical properties. A renormalization-group study\cite{Nienhuisfcb} of the face-cubic model 
goes back to 80th of last century. 
In two dimensions, the critical EFC model belongs to the same universality class
of the O($n$) loop model\cite{loopsquare} for $n<2$. For $n=2$,
the model belongs to an universality class that is different to the O(2) loop model due to the marginally
relevant cubic field\cite{GuoERC, dingERC}. For $n>2$, the phase 
transition of the model becomes discontinuous. 
In three-dimensional simple cubic lattice, the phase transition of the EFC model is continuous and belongs to 
the O($n$) loop universality class when $n\leq n_o$ with $n_o=2.1(1)$.
 For $n_o<n\leq n_t$ with $n_t=2.7(1)$, it still undergoes a second-order transition that is in a different 
universality class. For $n_t < n <n_p$ with $n_p\approx 4.54$, it displays a first-order phase transition at finite 
temperature $v<1$\cite{xuxiao}.

In two dimensions, a two-to-one Ising-spin representation of the configuration of the model is possible. Concretely, 
the Ising spins are located in the faces of the lattice (or the vertices of the dual lattice), and two nearest-neighboring faces should take different 
signs if and only if the edge between them is occupied by a bond.
Basing on the Ising-spin configurations, one can 
define magnetization and its Binder ratio in order to investigate the critical properties  of the model\cite{dingERC}.
Furthermore the critical properties of the model can also be captured by the percolation of the Ising clusters\cite{dingERC}.
In three dimensions, the Ising-spin representation is not applicable because of the different topology of space, 
but the critical properties can still be captured by the percolation of the clusters of the Eulerian graph\cite{xuxiao}.  

In current paper, we study the EFC model in a region with bond weight $v>1$. We find an 
Ising-like phase transition for all $n$. In two-dimensional square lattice, such a phase transition
is in fact an antiferromagnetic Ising phase transition. Its critical properties can be investigated 
by sampling a staggered magnetization and the corresponding Binder ratio, based on the Ising-spin representation. 
We further show that the critical properties of such a phase transition can also be 
captured by the percolation of the complement of the Eulerian graph.  
This percolation game is repeated in three-dimensional simple cubic lattice,
and a phase transition belongs to the three-dimensional Ising universality class is found.

\section{Algorithm}
For Monte Carlo simulation of the model, there are two efficient non-local algorithms can be selected. 
A cluster algorithm is given by Deng. {\it et. al}\cite{dengalg}, which is also applicable for O($n$) 
loop model. Such an algorithm is based on the Ising-spin representation of the configurations of the model, 
therefore it seems only applicable in two dimensions. In addition, for a two-dimensional lattice with periodic boundary conditions, 
a configuration with a single loop wrapping the system can't be represented by the Ising-spin configuration. 
This is not a severe problem because the critical property is not affected by the boundary conditions. 
However, if one want to study the model in the full state space, the worm algorithm\cite{worm} is a good choice. 
In three dimensions, because the Ising-spin representation is not applicable, the worm algorithm should be chosen.
Local algorithms, e.g., the plaquette update\cite{Winter}, can also be used,
but the simulation will severely suffer from critical slowing down.

The worm algorithm for the O($n$) loop model\cite{loop3D} or the EFC model can be combined
with the coloring trick\cite{color1,color2} in order to avoid 
a connectivity-checking procedure, which is a nonlocal procedure for a basic updating step thus is very time-consuming.
However, the coloring trick is applicable only for $n\ge1$, and the algorithm with coloring trick will obviously suffer from 
the critical slowing down for large $n$. Furthermore we find that, for a worm algorithm with coloring trick, 
the frozen part of the clusters often makes it very difficult for a worm returning to the start point
when the bond weight is much larger than 1, 
which leads to a  sharp reduce of the efficiency. Such problem is severe even for $n$ close to 1. 
Therefore, in our simulations, the connectivity checking is still necessary. Furthermore, one can use 
the simultaneous breadth-first connectivity checking\cite{breadth-first} to improve the efficiency.

\section{Variables and finite-size scaling formulae}
In two dimensions, the sampled variables are in two types. The first type is based on the Ising-spin configurations.
They are the staggered magnetization $m$ and its Binder ratio $Q$, 
which are defined as
\begin{eqnarray}
m&=&\langle \mathcal{M}\rangle,\\
Q&=&\frac{\langle\mathcal{M}^2\rangle^2}{\langle\mathcal{M}^4\rangle},
\end{eqnarray}
where $\mathcal{M}$ is 
\begin{eqnarray}
\mathcal{M}=\frac{\bigg|\sum\limits_{ia=1}^{N/2}S_{ia}-\sum\limits_{ib=1}^{N/2}S_{ib}\bigg|}{N},
\end{eqnarray}
with $S_{ia}$ ($S_{ib}$) the Ising spin in the $i$-th face of the sublattice $a$ ($b$), and $N$ 
the number of total faces. Such a definition of $m$
is similar to that of antiferromagnetic Potts model\cite{AFPotts3D}.
 
By sampling the two variables, we expect to determine the critical point $v_c$ and two critical exponent $y_t$ and $y_h$ by the following
finite-size scaling formulae\cite{Fss1,Fss2}
\begin{eqnarray}
Q&=&Q_0+a_1(v-v_c)L^{y_t}+a_2(v-v_c)^2L^{2y_t}+\cdots\nonumber\\
   &&+b_1L^{y_1}+b_2L^{y_2}+\cdots, \label{Qfss}\\
m&=&L^{y_h-d}(a+b_1L^{y_1}+b_2L^{y_2}+\cdots), \label{mfss}
\end{eqnarray}
where $L$ is the linear size of the system, and $d$ is the dimension of the lattice. 
$a_1$, $a_2$, $b_1$, and $b_2$ are unknown parameters, $y_1$ and $y_2$ are the correction-to-scaling exponents, which take negative values.
(\ref{mfss}) is valid only at the critical point.

The second type variables include the wrapping probability $R$ \cite{wrap} and the worm 
return time $\tau$, which can also be used to determine the critical point and the two critical exponents
by the following finite-size scaling formulae
\begin{eqnarray}
R&=&R_0+a_1(v-v_c)L^{y_t}+a_2(v-v_c)^2L^{2y_t}+\cdots\nonumber\\
  &&+b_1L^{y_1}+b_2L^{y_2}+\cdots, \label{Rfss}\\
\tau&=&L^{2y_h-d}(a+b_1L^{y_1}+b_2L^{y_2}+\cdots),\label{taufss}
\end{eqnarray}
where (\ref{taufss}) is valid only at the critical point.

The wrapping probability is defined as the probability that there exists a cluster that spans the system and 
connects itself along at least one of the directions, which forms a nontrivial loop that wraps the system. 
Obviously, such a definition is only applicable for a 
system with periodic boundary conditions. Generally, $R$ can be written as
\begin{eqnarray}
R=\langle \sum\limits_\alpha\mathcal{R}_\alpha\rangle/n_d
\end{eqnarray}
where $n_d$ is the number of directions. $\mathcal{R}_\alpha$ is 1 (0) if there is a (no) cluster wraps the system along the $\alpha$ direction,
 whether or not the cluster wraps along the other directions. 
For example, on the simple cubic lattice, $\alpha$ usually takes the Euclidean coordinate directions, thus $\alpha=x,y$, or $z$ and $n_d=3$.

The worm return time $\tau$ is defined as the number of updating attempts of the worm (it is larger than the length of the worm), 
which is proven to have the similar critical behavior as the susceptibility\cite{loop3D}. 

In saying percolation, one should distinguish the `complementary percolation' from the `normal percolation'. 
In the EFC model, each edge of the lattice has two states, linked or vacant. From now on, an edge 
is considered to be occupied by a `normal bond' if the state is linked, otherwise it is considered 
to be occupied by a `complementary bond'. The percolation on the lattice can be defined basing on the `normal bonds' 
or the `complementary bonds', which we call as `normal percolation' or `complementary percolation' respectively.
For the normal percolation of the EFC model, it has been studied both in two and three dimensions\cite{dingERC,xuxiao},
in current paper, we mainly pay attention to the complementary percolation.

\section{Results}
\subsection{Results on the square lattice}
\begin{figure}[htbp]
\includegraphics[scale=0.9]{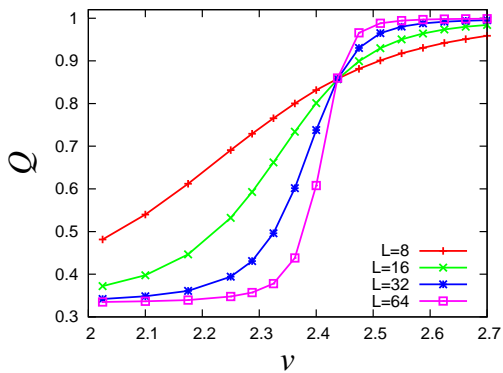}
\caption{(Color online) Plot of the Binder ratio $Q$ versus the bond weight $v$ 
for the $n=1.5$ EFC model on the square lattice. The systems shown here,
 with relatively small sizes, are just for illustrative purpose.}
\label{Q}
\end{figure}

\begin{figure}[htbp]
\includegraphics[scale=0.9]{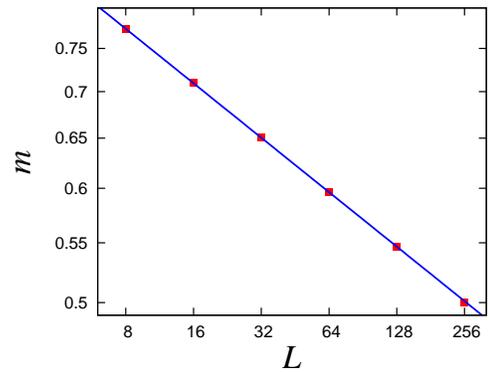}
\caption{(Color online) Log-Log plot of the staggered magnetization $m$ versus system size $L$ for the $n=1.5$ 
EFC model on the square lattice. The straight line, with slope -1/8, is added to guide eyes.}
\label{m}
\end{figure}

In two dimensions, our simulations are mainly performed on the square lattice. 
Our numerical procedure is illustrated by taking the $n=1.5$ EFC model as an example.
The largest system size that we reached is $L=256$ and each data point is obtained by the
average of $3\times 10^6 \sim 1\times 10^7$ samples. The error bar is computed by 
dividing the data into $1000\sim 2000$ bins.
 Fig. \ref{Q} is an illustrative plot of $Q$ versus $v$ for various system sizes.
The behavior of $Q$ obviously indicates a phase transition at the critical point $v_c\approx2.44$. 
A fit of the data near the critical point according 
to (\ref{Qfss}) gives $v_c=2.43684(2)$ and $y_t=0.997(5)$. 
The estimation of $y_t$ coincides with the exact value $y_t=1$ of the Ising model\cite{exact}, 
thus we expected the phase transition belongs to 
the Ising universality class. This is confirmed by the numerical estimation of $y_h$. 
Extensive simulations are done at the critical point $v_c$. The data of $m$ are shown in Fig. \ref{m}.
We fit the data according to (\ref{mfss}) and obtain $y_h=1.8750(4)$, which is in good agreement with the
exact result $y_h=15/8$\cite{exact}.

\begin{figure}
\includegraphics[scale=0.9]{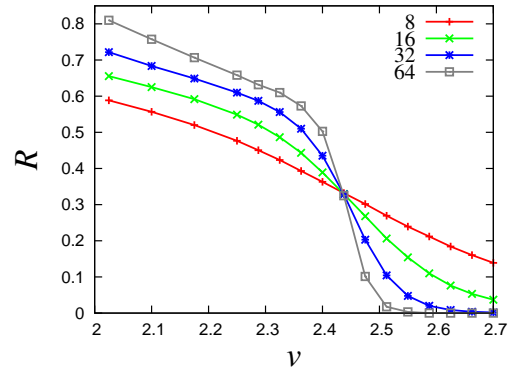}
\caption{(Color online) Plot of the wrapping probability $R$ (for the complementary percolation)
 versus the bond weight $v$ for the $n=1.5$ EFC model on the square lattice.
The systems shown here, with relatively small sizes, are just for illustrative purpose.}
\label{R}
\end{figure}

\begin{figure}[htbp]
\includegraphics[scale=0.9]{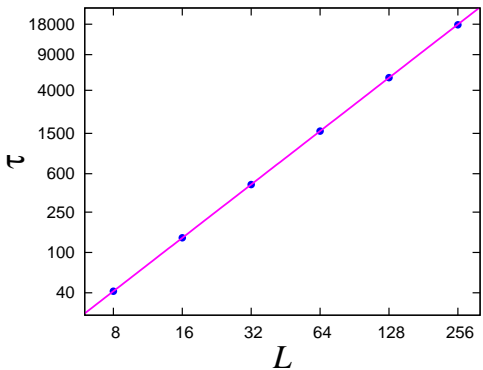}
\caption{(Color online) Log-Log plot of the worm return time $\tau$ 
 versus system size $L$ for the $n=1.5$ EFC model on the square lattice.
 The straight line, with slope 7/4, is added to guide eyes.}
\label{tau}
\end{figure}

Such an Ising-like phase transition can also be described by the complementary percolation of the Eulerian graph. 
Fig. \ref{R} is the plot of $R$ versus $v$, which can be compared with the plot of $Q$. 
By fitting the data near the critical point according to (\ref{Rfss}), we obtain $v_c=2.43685(3)$ and $y_t=1.02(3)$. 
These results are in agreement with the fit of $Q$.
Furthermore, the critical exponent $y_h$ can also be obtained from the fit of $\tau$ according to (\ref{taufss}), 
which gives $y_h=1.8756(7)$. It coincides with the result fit from $m$. 
Fig. \ref{tau} is an illustrative plot of $\tau$ versus system size $L$, with $v=v_c$. 

We also simulate the cases $n=0.5$, 1, 2, 2.5, 3, and 5. All results are listed in Table \ref{squaretab}.
The $y_t$ can be fit by (\ref{Qfss}) or (\ref{Rfss}) and $y_h$ can be fit by (\ref{mfss}) or (\ref{taufss}), 
in the table we list the best estimations.
It is clear that the critical exponents for all $n$ coincide with the exact values of the Ising model, namely the phase 
transition is in the universality class of the two-dimensional Ising model.

\begin{table}[htbp]
\caption{Critical properties of the Ising-like transition of the EFC model on 
the square lattice, the estimations of $y_t$ and $y_h$ can be compared with the exact results: $y_t=1$ and $y_h=15/8$\cite{exact}.}
 \begin{tabular}{c|l|l|l}
    \hline
      $n$   &$v_c$     &$y_t$    &$y_h$ \\
    \hline
    $0.5$   &2.3932(1)  &1.00(1) &1.876(1) \\
    $1.0$   &2.41421(1) &1.001(3) &1.8749(2) \\
    $1.5$   &2.43684(2) &0.997(5) &1.8750(4) \\
    $2.0$   &2.46100(3) &0.99(1)  &1.875(1)  \\
    $2.5$   &2.48663(3) &0.99(1)  &1.8748(5)  \\
    $3.0$   &2.51409(6) &0.99(2)  &1.876(2) \\
    $5.0$   &2.6419(1)  &0.98(3)  &1.874(2)\\
    \hline
\end{tabular}
\label{squaretab}
\end{table}

\subsection{Results on the simple cubic lattice}
In three dimensions, our simulations are performed on the simple cubic lattice. 
The sampled variables are the wrapping probability for the complementary percolation and 
the worm return time. The staggered magnetization and its Binder ratio are no longer applicable.
The largest system size that we reached is generally $L=48$ and the finite-size scaling analysis is
similar to that in two dimensions. All the results are listed in Table \ref{cubictab}. 
For the critical exponents of the three-dimensional Ising model, there is no exact result,
but there are numerical ones for comparison\cite{loop3D,henkIsing3D,youjinIsing3D,xuxiao}. 
Our estimations of $y_t$ and $y_h$ are consistent with the three-dimensional Ising model\cite{henkIsing3D,youjinIsing3D}.
Therefore the phase transition is believed to be in the universality class of the three-dimensional Ising model.

\begin{table}[htbp]
\caption{Critical properties of the Ising-like transition of the EFC model on 
the simple cubic lattice, the estimations of $y_t$ and $y_h$ can be compared with the Monte Carlo results 
$y_t=1.5868(3)$ and $y_h=2.4816(1)$\cite{youjinIsing3D}, which are much more accurate.}
 \begin{tabular}{c|c|c|c}
    \hline
      $n$   &$v_c$     &$y_t$    &$y_h$\\
    \hline
    $1.0$   &4.58516(3) &1.58(1) &2.482(3) \\
    $1.5$   &4.58526(5) &1.59(2) &2.484(5) \\
    $5.0$   &4.58565(5) &1.57(3) &2.483(5) \\
    \hline
\end{tabular}
\label{cubictab}
\end{table}

\section{The nature of the phase transition}
In conclusion, we have found a critical line of the $n$-component EFC model 
on the square lattice and the simple cubic lattice in the region $v>1$. The phase 
transition belongs to the Ising universality class for all $n$. 
Here we discuss the nature of this phase transition.

For the face-cubic model in (\ref{hfcb}) with a given value of $n$, as
the temperature $1/K$ decreases, the system undergoes a second-order phase
transition at a finite temperature $1/K_c>0$, below which the face-cubic 
symmetry is spontaneously broken and a long-range order develops for $n\le 2$ 
in two dimensions, although the transition belongs to the O($n$) universality class and 
the cubic anisotropy is irrelevant for $n<2$ \cite{GuoERC, LTOn}. 
In the language of the graphical model (\ref{fcb}), an infinite cluster
emerges at the critical point $0<v^{(\rm{cubic})}_c=\tanh K_c<1$, which is 
actually the (normal) percolation threshold of clusters.
For $n>2$, the transition line continues in the graphic model (\ref{fcb}) 
 while the corresponding coupling $K_c$ lies outside physical 
region\cite{dengalg}. The cubic symmetry is expected to be broken when $v>v^{(\rm{cubic})}_c$.

\begin{figure}[htbp]
\includegraphics[scale=0.9]{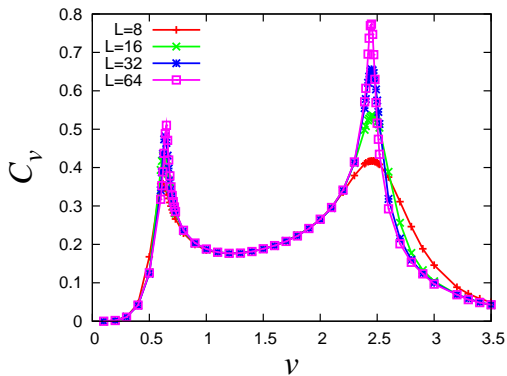}
\caption{(Color online) Plot of the fluctuation of bonds $C_v$ versus the bond weight $v$ for the
 $n=1.5$ EFC model on the square lattice.}
\label{Cv}
\end{figure}

\begin{figure}[htbp]
\includegraphics[scale=0.9]{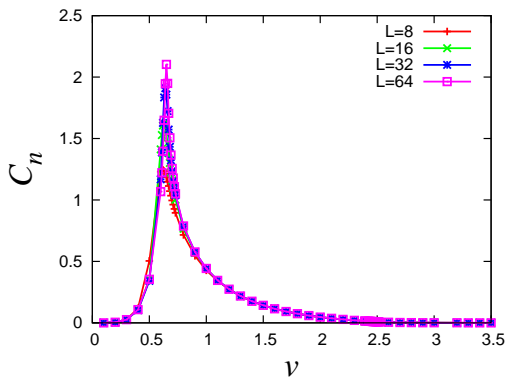}
\caption{(Color online) Plot of the fluctuation of clusters $C_n$ versus the bond weight $v$ for the
 $n=1.5$ EFC model on the square lattice.}
\label{Cn}
\end{figure}

The Ising-like phase transition occurs in the region $v>v^{(\rm{cubic})}_c$. 
Since the cubic symmetry of the model has been broken for all $n$, 
it suggests that $n$ no longer takes effect in the Ising-like transition. 
To explore such an argument, 
we measure the specific-heat-like quantities, the fluctuation of bonds $C_v=(\langle N_b^2\rangle -\langle N_b\rangle^2)/L^2$ 
and the fluctuation of clusters $C_n=(\langle N_c^2\rangle-\langle N_c\rangle^2)/L^2$,
which are plotted in Fig. \ref{Cv} and Fig. \ref{Cn} respectively.
It is shown that there are two peaks of $C_v$ but only one peak
of $C_n$. The peak of $C_v$ in the region $v<1$ corresponds to the threshold of the 
normal percolation of the Eulerian graph, while the second peak corresponds to the 
threshold of the complementary percolation.
The appearance of the peak of $C_n$ at the threshold of the normal percolation means 
the transition depends on $n$, which interprets the fact that 
the critical exponents vary with $n$ at this point\cite{dingERC}. 
The lack of second peak of $C_n$ at the threshold of the complementary percolation indicates that 
the transition is independent of parameter $n$. This is also reflected by the fact that, by varying parameter $n$, the
critical point $v_c$ is only shifted by a small value, as shown in Tables I and II.


Since the properties of the transition are independent of $n$ in the region 
$v>v^{(\rm cubic)}_c$, 
the universality class of the phase transition can be determined by an arbitrary $n$, it is convenient to chose $n=1$. 
On an Eulerian lattice (coordinate number is even) such as the square lattice or the simple cubic lattice,
a complementary graph of the EFC model is also an Eulerian one. This leads to a result that
the complement of the $n=1$ EFC model is also an $n=1$ EFC model but with bond weight $1/v$. Therefore the 
complementary percolation transition of the Eulerian graph must belong to the Ising universality class.
Such a mapping also allows us to predict the critical point of the complementary percolation 
of the $n=1$ EFC model on an Eulerian lattice as 
\begin{equation}
v_c({\rm complementary})=1/v_c({\rm normal}).\label{com_map}
\end{equation}
On the square lattice, the threshold of the normal percolation of the $n=1$ EFC model is $v_c=\sqrt{2}-1$\cite{dingERC},
 it gives the threshold of the complementary percolation $v_c=\sqrt{2}+1$, which coincides with our numerical result
$v_c=2.41421(1)$ (Table \ref{squaretab}). 
On the simple cubic lattice, the percolation threshold of the normal percolation of the 
$n=1$ EFC model is $v_c=\tanh K_c=0.2180944(1)$\cite{henkIsing3D}. 
This predicts the critical point of the complementary percolation $v_c=4.585170(1)$, which is also verified by our 
numerical result (Table \ref{cubictab}).

In two dimensions, the Ising-like phase transition can be understood via the dual-spin representation of the EFC
model. For the square lattice, the symmetry of the 
odd-even sites in the dual lattice is broken when the cluster becomes 
dense enough at very large $v$, which leads to an AF pattern in the dual-spin configurations. 
 An Ising-like transition from a disorder phase to
an AF ordered phase thus presents independent of $n$. 
The AF pattern obviously depends on the lattice structure, thus the Ising-like phase transition 
is lattice-dependent.
For example on the honeycomb lattice in which the EFC model is equivalent to the
O($n$) loop model due to the coordination number 3, we can't find such an Ising-like 
phase transition. 
There is only one peak of $C_n$ and the threshold of the complementary percolation coincides
with that of the normal percolation, i.e., the critical point $v_c=1/\sqrt{2+\sqrt{2-n}}$\cite{Nienhuisexact}.
For the simple cubic lattice, the simple dual-spin picture is
not applicable. We do not have a simple physical picture to interpret the
Ising-like behavior.  It is not easy to predict the existence of such Ising-like phase 
transition in a three-dimensional lattice. It must be studied case by case,
 especially for the lattice that is not Eulerian.

It is interesting to compare this Ising transition with the 
recently studied Ising-like
transitions in the square O($n$) loop model \cite{FGB}, in which the critical 
behaviors in the region $n>2$ and $n<2$ are different, although the critical 
lines connect at $n=2$. In the spin language of the two models, the 
low-temperature phase of the face-cubic model is ordered for all $n$, but that 
of the loop model is critical in the sense that the spin-spin correlation 
decays algebraically for $n \le 2$, and the O($n$) critical line ends at $n=2$.
As a result of this fact, the universality class of the Ising-like transition
in the O($n$) loop model
is a superposition of the low-temperature O($n$) critical behavior and the Ising
behavior\cite{FGB} for $n <2$, while for $n>2$ the transition resembles a hard-square lattice gas transition.

\section*{Acknowledgment}
This work is supported by the National Science Foundation of China
 (NSFC) under Grant Nos. 11205005 (C.D.), 11175018 (W.G.), and 11275185 (Y.D.).

\end{document}